\documentclass{jps-cp}
\usepackage{txfonts} %Please comment out this line unless the txfonts package is availabe in your LaTeX system.
\usepackage{siunitx}
\usepackage{lineno}

\title{The CLICTD Monolithic CMOS Sensor}

\author{Katharina \textsc{Dort}$^{1,2}$ on behalf of the CLICdp collaboration}

\inst{$^{1}$ European Organization for Nuclear Research (CERN), Geneva, Switzerland \\
$^{2}$ II Physics Institute, University of Giessen, Germany}

\email{katharina.dort@cern.ch}

\recdate{November 16, 2020}

\abst{

CLICTD is a monolithic silicon pixel sensor fabricated in a modified 180 nm CMOS imaging process with a small collection electrode design and a high-resistivity epitaxial layer. It features an innovative sub-pixel segmentation scheme and is optimised for fast charge collection and high spatial resolution. The sensor was developed to target the requirements for the tracking detector of the proposed future Compact Linear Collider (CLIC). Most notably, a temporal resolution of a few nanoseconds and a spatial resolution below \SI{7}{\micro m} are demanded. In this contribution, the sensor performance measured in beam tests is presented with emphasis on recent studies using assemblies with different thicknesses (down to \SI{50}{\micro m} to minimize the material budget) and inclined particle tracks.

}

\kword{Particle tracking detectors (solid-state detectors), Monolithic silicon pixel detectors, CMOS sensors with a small collection diode}

\begin{document}
\maketitle
%\linenumbers
\section{Introduction}

The CLIC tracker detector (CLICTD) is a pixelated monolithic CMOS sensor featuring a high resistivity epitaxial layer and a small collection electrode for reduced sensor capacitance. 
CLICTD targets the requirements of next-generation tracking detectors developed for future Higgs factories such as the Compact Linear Collider (CLIC).
CLIC is a concept for a linear electron-positron collider with centre-of-mass energies between 380\,GeV and 3\,TeV.
The physics program comprises Standard Model top quark and precision Higgs physics as well as searches for Beyond Standard Model physics~\cite{CLIC_2018_Summary}. 

Physics-driven requirements and experimental conditions impose stringent requirements on the CLIC detector systems~\cite{Dannheim:2673779}.
For the tracking detector, a spatial resolution of $< \SI{7}{\micro m}$ in the direction perpendicular to the magnetic field and a maximum material budget of 1--\SI{2}{\percent}$X_0$ are needed to achieve the required measurement accuracy. 
Additionally, a hit time tagging resolution of 5\,ns is required to suppress out-of-time beam-induced background particles. 
An average power dissipation of $150 \textrm{mW/cm}^2$ should not be exceeded to allow for a low-mass leak-less water cooling system.
Finally, the hit detection efficiency should surpass \SI{99.7}{\percent}.

Monolithic CMOS silicon sensors are considered promising candidates for the large-area tracking detector due to their low material budget and large-scale production capabilities. 
In particular, monolithic sensors with a small collection diode have shown to profit from a low detection threshold and a high signal-to-noise ratio~\cite{AglieriRinella:2017lym}.  
Sensor processes optimised for prompt charge collection have been designed~\cite{Munker_2019} and successfully tested with regard to radiation hardness~\cite{Dyndal:2019nxt}, high efficiency and fast charge collection~\cite{proceedings_ichep}. 

In this document, the performance of the monolithic small collection diode CMOS sensor CLICTD is presented. 
Most notably, the effect of different sensor bias voltages on the performance is investigated and the active depth of the sensor is studied with inclined particle tracks.
Moreover, the performance of CLICTD assemblies thinned down to \SI{50}{\micro m} is evaluated. 

\section{The CLICTD Monolithic Sensor}
\label{sec:clictd}
CLICTD is a pixelated monolithic high-resistivity CMOS sensor with an active matrix of 16\,x\,128 detection channels, each measuring \SI{300x30}{\micro m}.
In the following, the sensor process and the front-end design are outlined.

\subsection{Sensor process}

\begin{figure}[!b]
	\centering
	\begin{minipage}[b]{0.49\textwidth}
		\includegraphics[width=\linewidth]{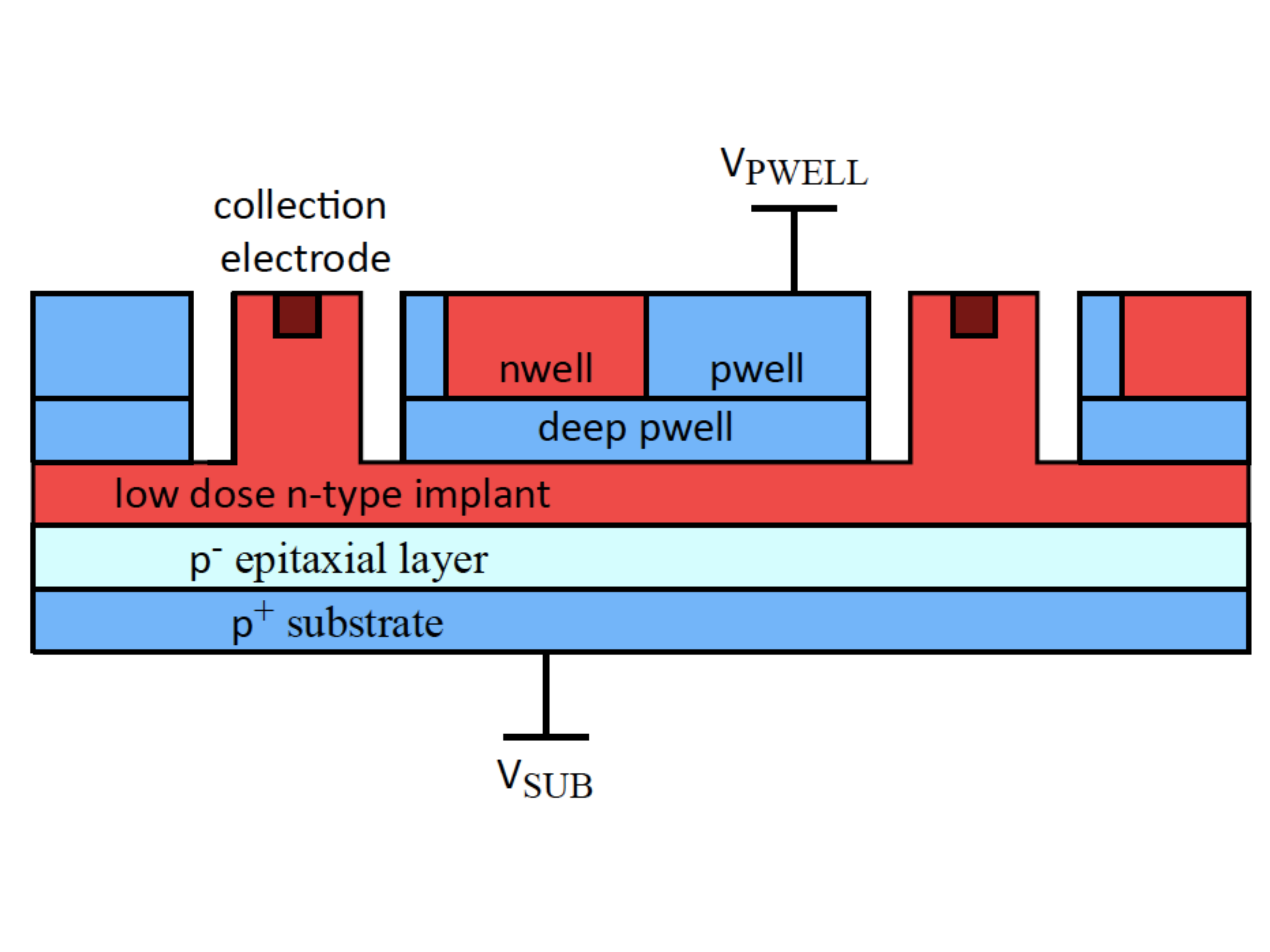}
		%\caption{Bla bla}
		%\label{fig:CLICTD_sensor_cont}
	\end{minipage}
	\hfill
	\begin{minipage}[b]{0.49\textwidth}
		\includegraphics[width=\linewidth]{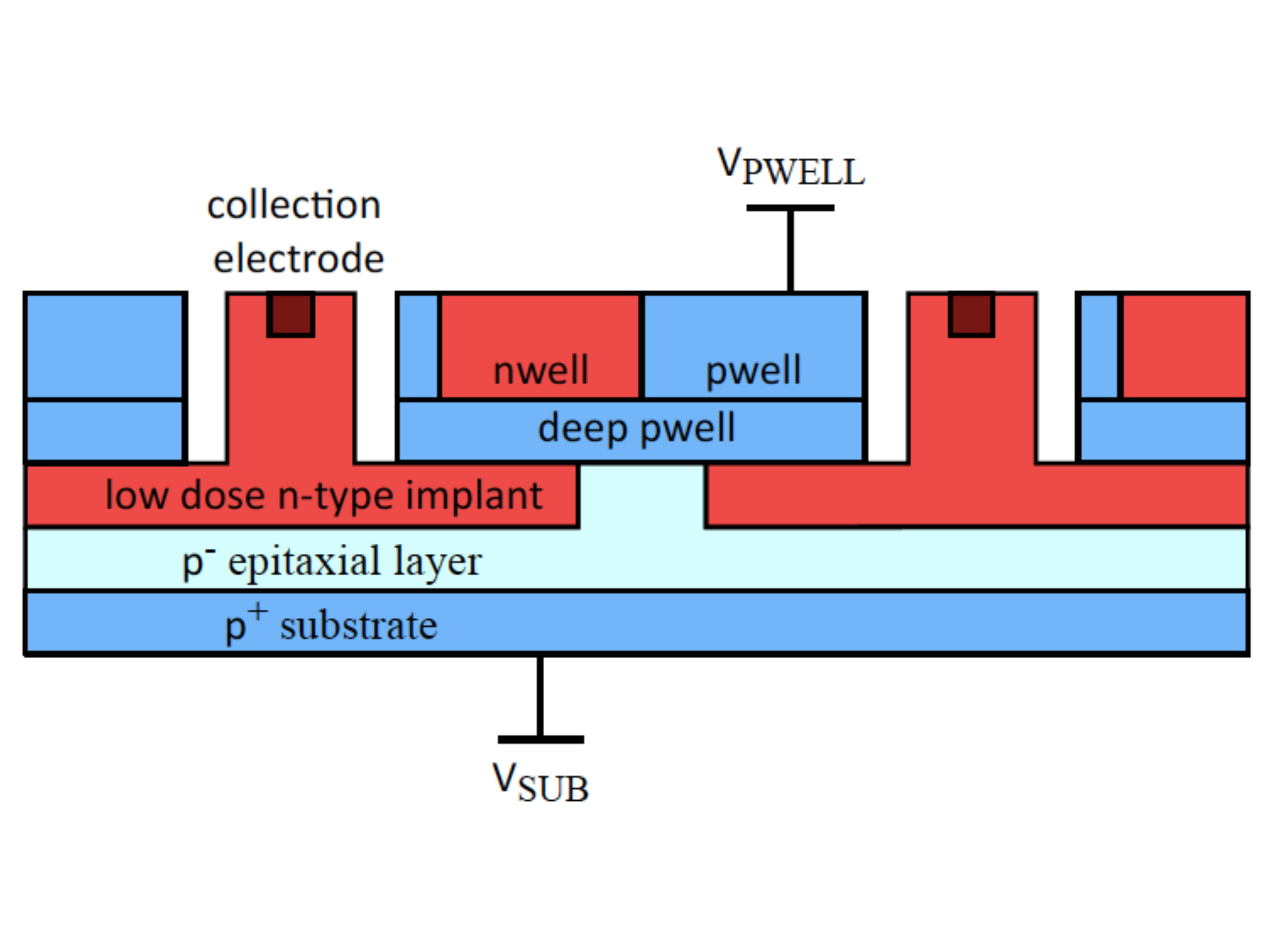}
		%\caption{Bla bla.}
		%\label{fig:CLICTD_sensor_gap}
	\end{minipage}
	\caption{CLICTD sensor process variants: continuous n-type implant (left) and segmented n-type implant (right)}
	\label{fig:CLICTD_sensors}
\end{figure}

CLICTD is fabricated in two different variants of a modified 180\,nm CMOS imaging process~\cite{SNOEYS201790}, which are shown in Fig.~\ref{fig:CLICTD_sensors}.
In both process variants, a small n-type collection electrode is placed on top of a high-resistivity p-type epitaxial layer with a thickness of \SI{30}{\micro m}.
The epitaxial layer is grown on top of a highly doped p-type backside substrate.
The analogue and digital on-channel circuitry is placed on deep p-wells to shield them from the electric field inside the sensor. 
Additionally, the sensor is shielded from the fast-switching circuitry that can act as  a noise source. 
The reverse bias voltage, which is limited to -6\,V, is applied to the substrate and the p-wells.
The applied bias voltage suffices to achieve full lateral depletion in the epitaxial layer due to the introduction of an n-type implant below the p-wells~\cite{SNOEYS201790}.
For lower absolute bias voltages, the sensor is not fully depleted around the small collection diode, which leads to a higher sensor capacitance~\cite{clictd_design_characterization}.

In a second process variant, the n-type implant is segmented, as illustrated on the right-hand side of Fig.~\ref{fig:CLICTD_sensors}.
The segmentation gives rise to an increase in the lateral electric field, which accelerates charge collection and suppresses the motion of charge carriers to neighbouring pixel cells (inhibited \textit{charge sharing})~\cite{Munker_2019}.
In CLICTD assemblies with segmented n-implant, the segmentation is only applied in the column direction since charge sharing in the row direction is desired to improve the spatial resolution.

\subsection{Analogue and digital front-end}

The elongated \SI{300x30}{\micro m} detector channels are segmented into eight sub-pixels along the \SI{300}{\micro m} dimension. 
Each sub-pixel is equipped with its own collection diode and  analogue front-end.
The eight sub-pixel outputs are combined with an $OR$ gate in the shared digital logic of a channel~\cite{clictd_design_characterization}. 
This  front-end scheme allows to save space for the digital circuitry without interfering with the small collection diode design. 

CLICTD is able to perform simultaneous timing and energy measurements.
For the timing measurement, the Time-of-Arrival (ToA) is recorded in 10\,ns bins.
For the energy measurement, the Time-over-Threshold (ToT), defined  as the number of clock cycles the signal is above the detection threshold, is recorded.
As a consequence of the combination of sub-pixel measurements in a channel, the ToA is set by the sub-pixel with the earliest time-stamp.
The ToT is set by the sub-pixel with the highest signal in case only one particle is measured or the total number of clock cycles during which at least one sub-pixel is above threshold  for multiple discrete hits.  
The binary sub-pixel hit pattern is recorded as well.

The propagation of charge carriers to adjacent sub-pixels can lead to the activation of multiple sub-pixels in a channel. 
Therefore, the reduction of charge sharing for the process with the gap in the n-layer is advantageous to suppress the response of more than one sub-pixel per channel.
%\section{Test-beam set-up}

\section{Performance in Particle Beams}

The performance of CLICTD has been studied at the DESY II test-beam facility~\cite{Diener:2018qap} using a EUDET-type  beam telescope~\cite{Jansen:2016bkd}, with an additional Timepix3 plane~\cite{Poikela} serving as a timing reference.
The offline analysis of the test-beam data is performed with the test-beam reconstruction framework Corryvreckan~\cite{corry_software}.
In the following sections, selected results with a focus on different reverse bias voltages are presented. 

The reverse bias voltage applied to CLICTD has an effect on the sensor as well as on the front-end performance: 
On the one hand, the operation of the front-end is more challenging at high absolute bias voltages due to a slow-down of the on-channel NMOS transistors~\cite{clictd_design_characterization}. 
On the other hand, the sensor performance improves for high absolute bias voltages, according to 3D TCAD simulation studies~\cite{Munker_2019}:
%best sensor performance is expected at the highest possible (absolute) bias voltage of -6\,V/-6\,V for substrate/p-wells~\cite{clictd_design_characterization}.
At lower bias voltages, the sensor area around the collection diode is not fully depleted leading to a higher sensor capacitance, as explained in Section~\ref{sec:clictd}.
A lower capacitance leads to reduced noise while enhancing the signal, which makes it possible to operate the sensor at a lower detection threshold. 
In laboratory studies, a minimum operation threshold of approximately $140$\,e$^-$ at the highest possible bias voltage $-6\,\textrm{V}/-6\,\textrm{V}$, for substrate/p-wells has been identified.
For a lower absolute bias voltage of $-3\,\textrm{V}/-3\,\textrm{V}$, the minimum threshold rises by approximately \SI{40}{\percent}  to 200\,e$^-$, indicating that the improvement of the sensor performance outweighs the limitations of the  front-end at high absolute bias voltages~\cite{clictd_design_characterization}.

\begin{figure}[btp]
	\centering
	\begin{minipage}[b]{0.49\textwidth}
		\includegraphics[width=\linewidth]{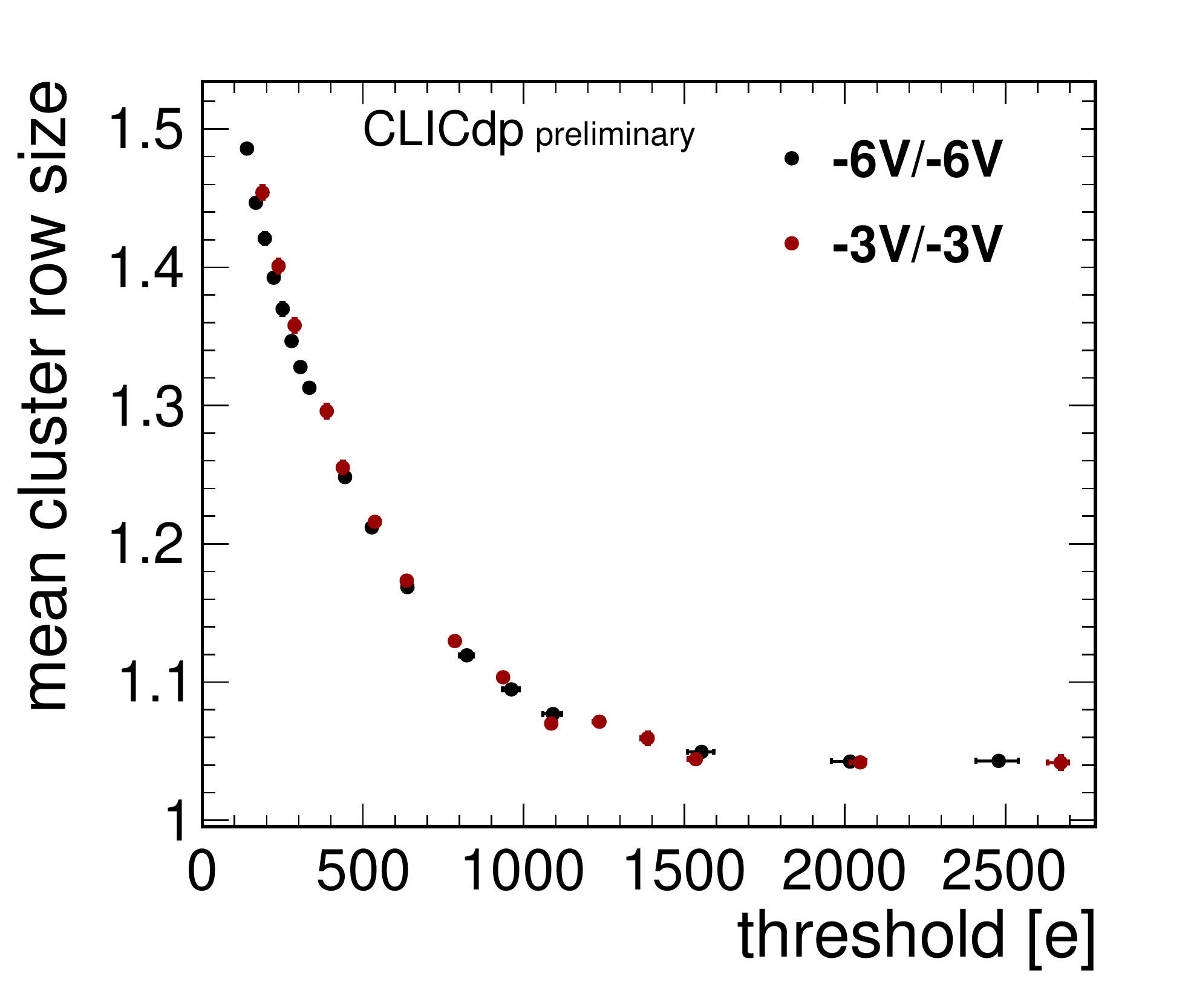}
		\caption{Mean cluster size in row direction for $-3\,\textrm{V}/-3\,\textrm{V}$ and $-6\,\textrm{V}/-6\,\textrm{V}$  applied to substrate and p-wells.}
		\label{fig:cls_size}
	\end{minipage}
	\hfill
	\begin{minipage}[b]{0.49\textwidth}
		\includegraphics[width=\linewidth]{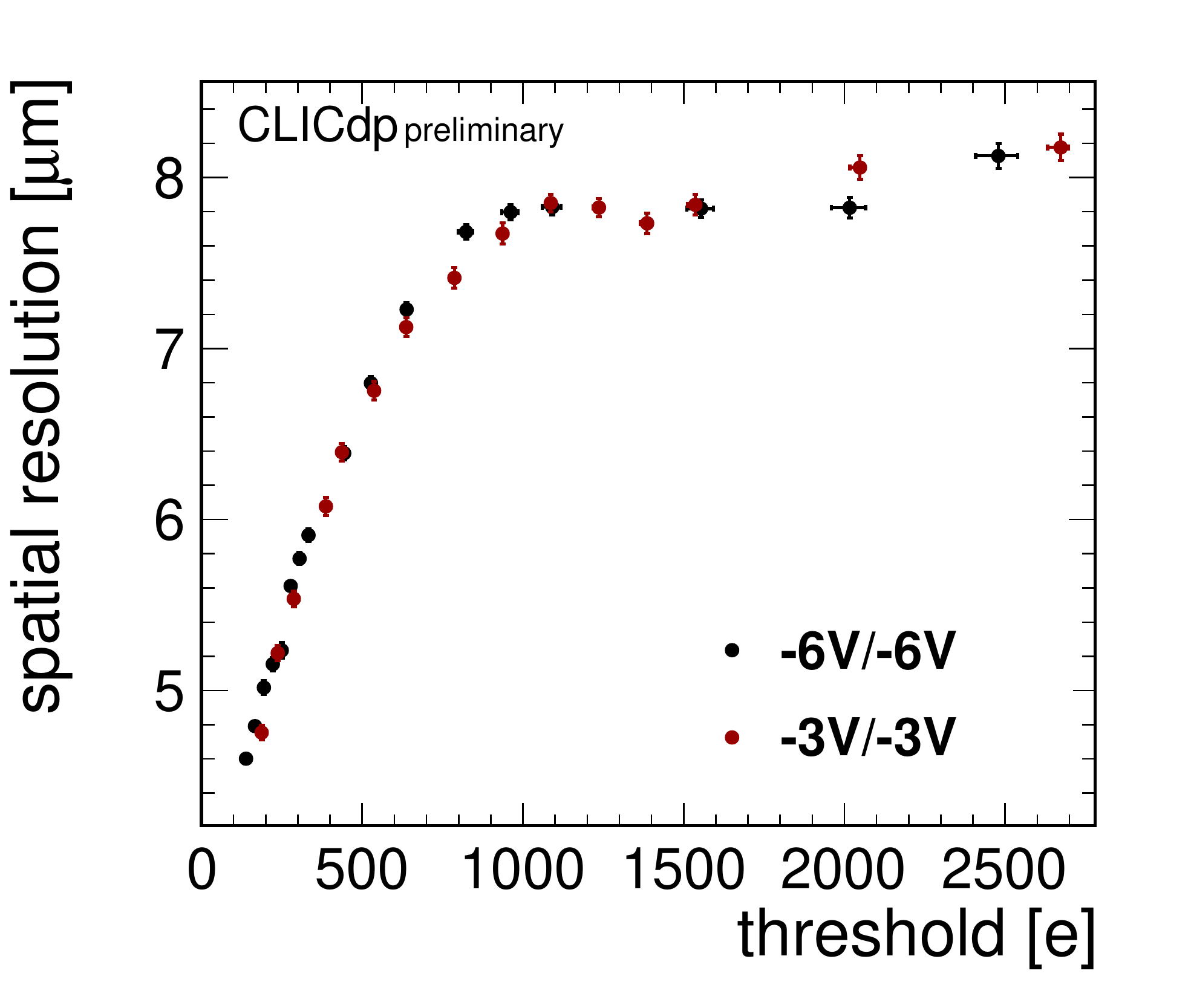}
		\caption{Spatial resolution (row direction) for $-3\,\textrm{V}/-3\,\textrm{V}$ and $-6\,\textrm{V}/-6\,\textrm{V}$  applied to substrate and p-wells.}
		\label{fig:spatial_res}
	\end{minipage}
	%\label{fig:rotation_size}
\end{figure}

The detection threshold has a critical impact on the performance of CLICTD, as illustrated in Fig.~\ref{fig:cls_size}, where the mean cluster size in row direction for an assembly with continuous n-layer is shown as a function of the threshold.
The mean row cluster size decreases rapidly for increasing thresholds before it starts to converge to a size of one. 
The initial sharp decline implies that a small change in threshold has an appreciable impact on the cluster size. 
While the cluster size for a given threshold is barely affected by the difference in bias voltage, the different operational  thresholds are responsible for the higher achievable cluster size at higher absolute bias voltages. 
The mean cluster size in row direction for $-6\,\textrm{V}/-6\,\textrm{V}$  is 1.49  at the minimum threshold of 140\,e$^-$ and 1.45 for $-3\,\textrm{V}/-3\,\textrm{V}$ at 200\,e$^-$.

The spatial resolution is strongly correlated with the cluster size, as shown in Fig.~\ref{fig:spatial_res} for the same assemblies.
A high cluster size improves the position reconstruction owing to charge weighted position interpolation. 
Consequently, the declining cluster size for higher thresholds is accompanied by a degradation in the spatial resolution. 
At the minimum threshold, the spatial resolution is \SI{4.6}{\micro m} for $-6\,\textrm{V}/-6\,\textrm{V}$  and  \SI{4.8}{\micro m} for $-3\,\textrm{V}/-3\,\textrm{V}$.

Further studies at $-6\,\textrm{V}/-6\,\textrm{V}$ have shown that a timing resolution of \SI{5.8}{ns} and an efficiency of $>$\SI{99.8}{\percent} can be achieved~\cite{proceedings_ichep}, which fulfil the performance requirements of the tracking detector. 
In the following sections, the bias voltage is fixed to $-6\,\textrm{V}/-6\,\textrm{V}$ .

%\begin{figure}[tbh]
%	\includegraphics[width=\linewidth]{images/5deg_75deg_overlay_sizeX.pdf}
%	\caption{bla bla}
%	\label{f1}
%\end{figure}

\section{Estimation of Active Depth}

\begin{figure}[btp]
	\centering
	\begin{minipage}[b]{0.49\textwidth}
		\includegraphics[width=\linewidth]{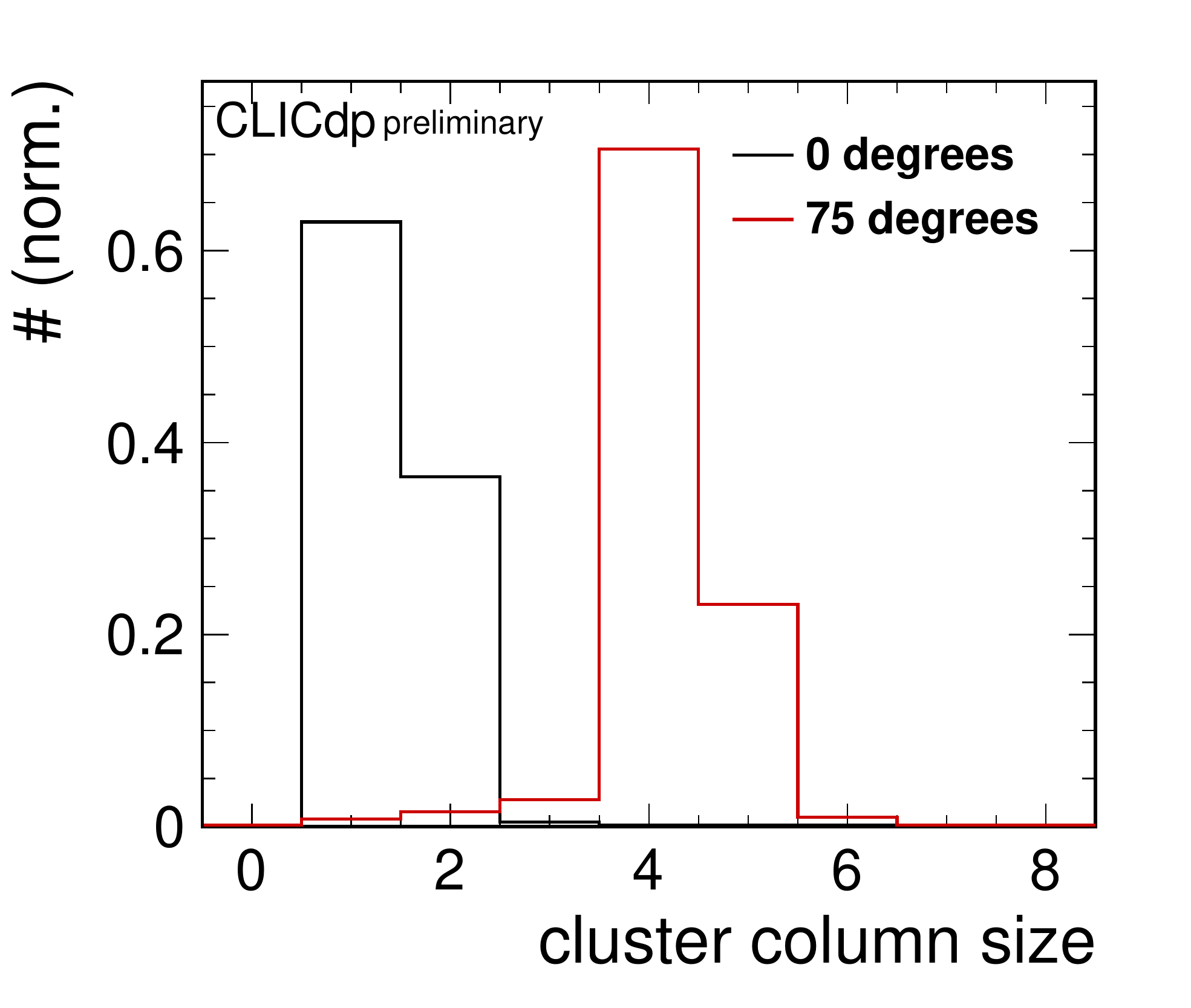}
		%\caption{Cluster size in column direction for 5 degree and 75 degree rotation angle. }
		%\label{fig:rotation_size_x}
	\end{minipage}
	\hfill
	\begin{minipage}[b]{0.49\textwidth}
		\includegraphics[width=\linewidth]{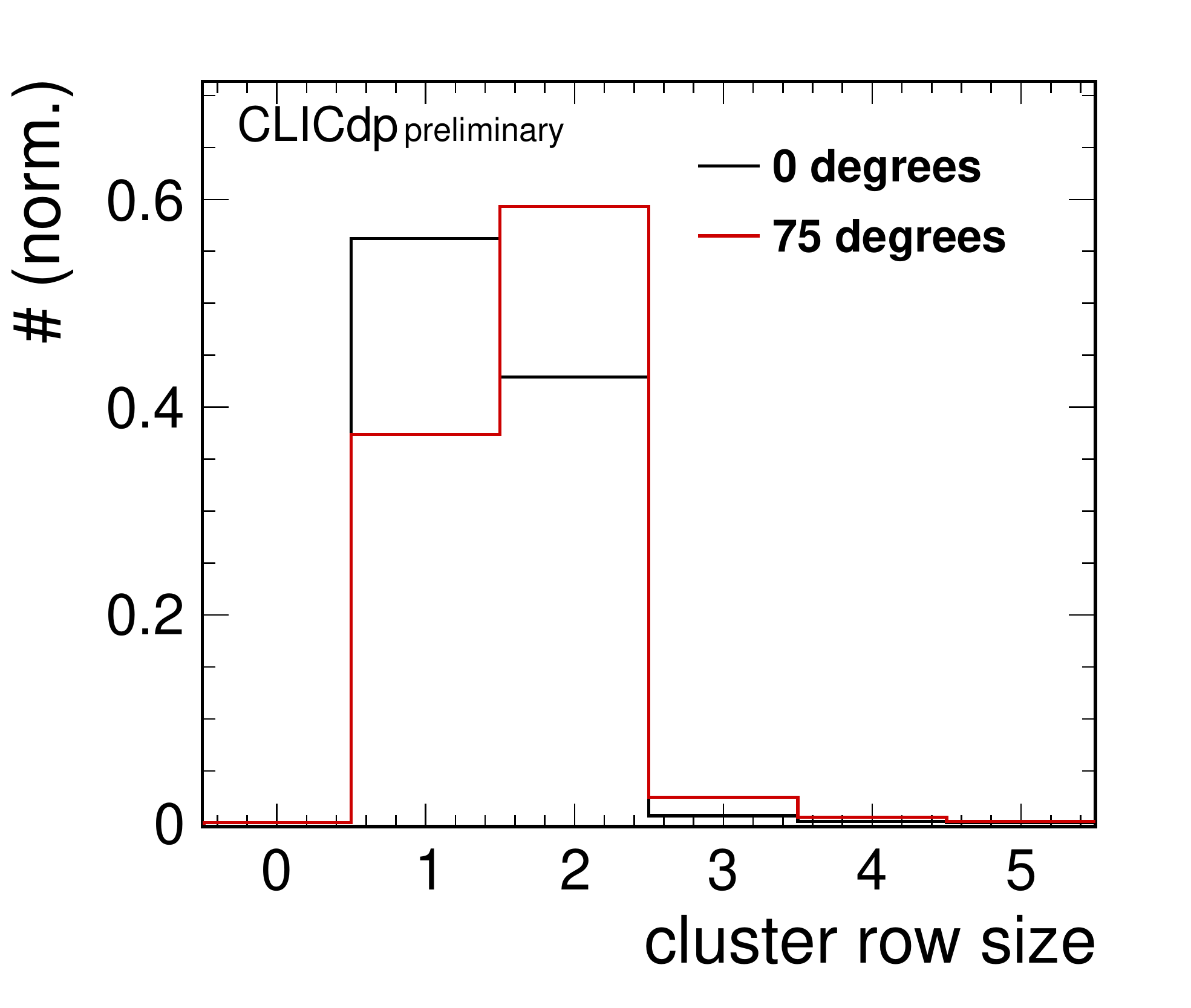}
		%\caption{Cluster size in row direction for 5 degree and 75 degree rotation angle.}
		%\label{fig:rotation_size_y}
	\end{minipage}
	\label{fig:rotation_size}
	\caption{Cluster column size (left) and cluster row size (right) for a rotation angle of $5^{\circ}$ and $75^{\circ}$. 
		The sensor is tilted in column direction.}
\end{figure}

%The charge collection as a function of the sensor depth is studied with the \textit{granzing angle technique}~\cite{Henrich:2002ed}.
The charge collection as a function of the sensor depth is studied with inclined particle tracks. 
By rotating the sensor in the particle beam, the amount of active silicon crossed by beam particles can be varied. 
For high rotation angles, particle tracks traverse several adjacent pixel cells, which gives rise to a larger cluster size, as can be observed in Fig.~\ref{fig:rotation_size}, where the cluster size distribution for two different rotation angles is shown for an assembly with continuous n-layer. 
On the left-hand side, the cluster size in the tilted direction is depicted.
The cluster size increases considerably since particles deposit energy in a line of multiple adjacent pixel cells.
In the perpendicular dimension, which is shown on the right, a small increase in cluster size can be seen. 
%even though the sensor is tilted in the other direction.
The increase can be explained by the higher total energy deposition that enhances charge sharing in both spatial dimensions. 

\begin{figure}[tbh]
	\centering
	\begin{minipage}[b]{0.49\textwidth}
		\includegraphics[width=\linewidth]{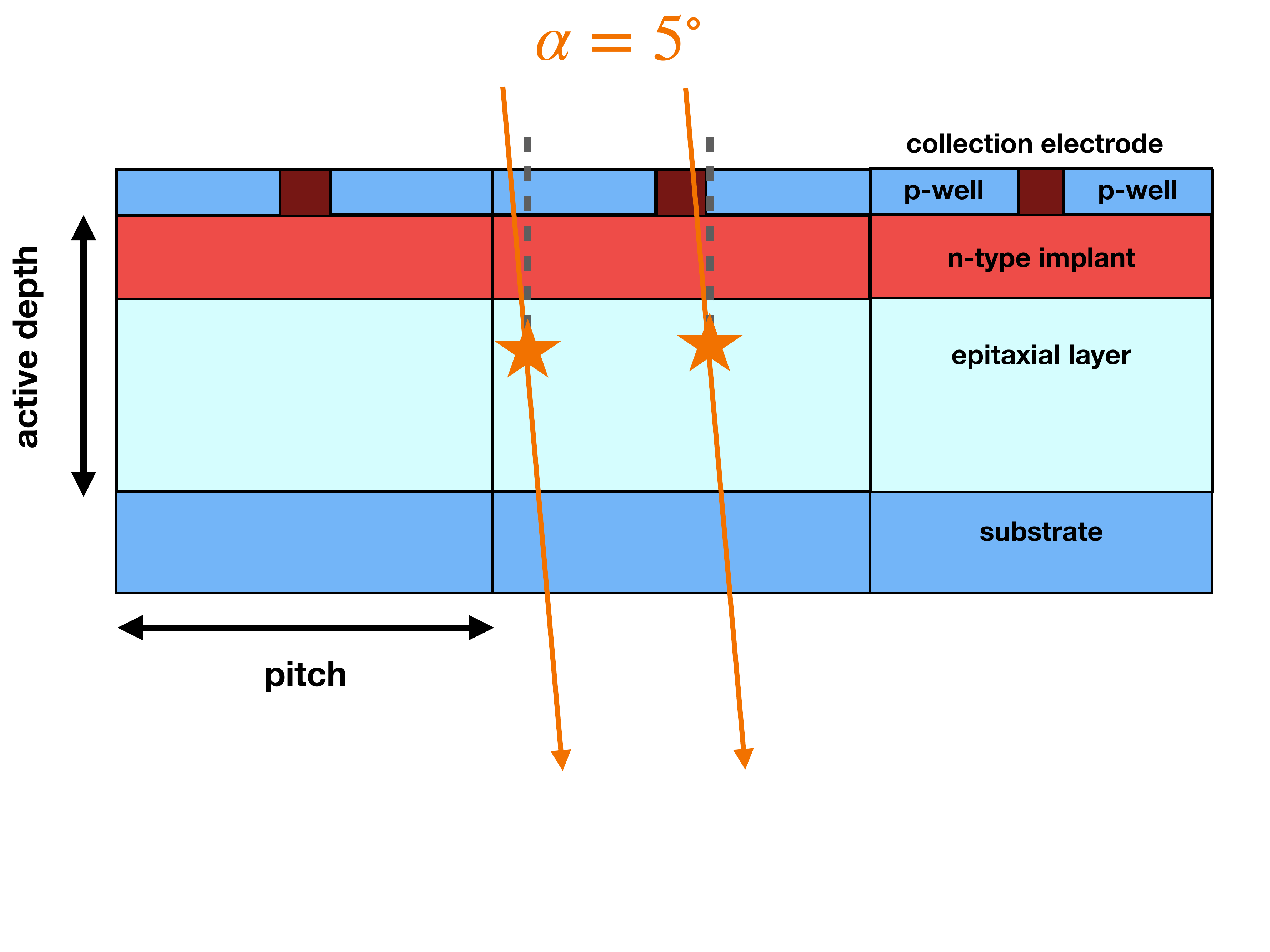}
		%\caption{Schematic  representation of the CLICTD sensor for two different track incident angles. }
		%\label{fig:sketch_geometry}
	\end{minipage}
	\hfill
	\begin{minipage}[b]{0.49\textwidth}
		\includegraphics[width=\linewidth]{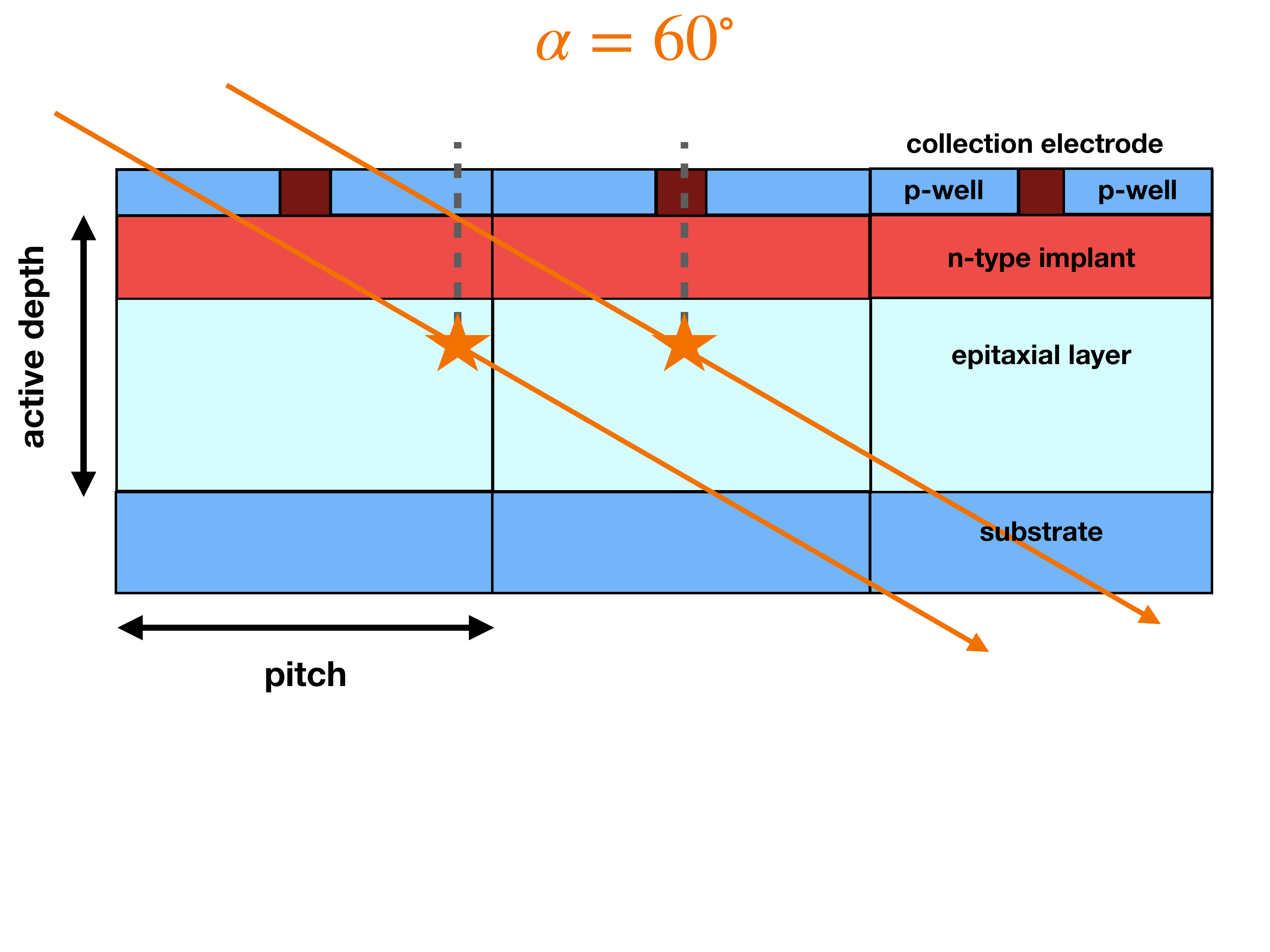}
		%\caption{Mean cluster size in column direction as a function of the rotation angle. }
		%\label{fig:active_depth_fit}
	\end{minipage}
	\caption{Schematic representation (not to scale) of inclined particle tracks impinging on pixel cells for $\alpha = 5^\circ$ (left) and $\alpha = 60^\circ$ (right). The stars indicate the reconstructed track intercept with the sensor.}
	\label{fig:rotation_sketches}
\end{figure}

From geometrical considerations, the cluster size in the spatial dimension that is titled can be related to the incident angle $\alpha$ and the \textit{active sensor depth} $d$, as sketched in Fig.~\ref{fig:rotation_sketches}. 
Charge carriers that are created below the active depth do not contribute to the measured signal. 
Therefore, the mean cluster size in the titled direction (column direction) can be expressed as: 
\begin{equation}
	\textrm{column cluster size} = \frac{d \tan{\alpha}}{\textrm{pitch}} + s_0
	\label{eq:active_depth}
\end{equation}
where the offset $s_0$ is the cluster column size for no rotation ($\alpha = 0$). 
This simple geometrical model neglects non-rotation induced charge sharing, i.e. charge sharing via pixel crosstalk or diffusion is not accounted for.

\begin{figure}[!tbp]
	\centering
	\begin{minipage}[b]{0.49\textwidth}
		\includegraphics[width=\linewidth]{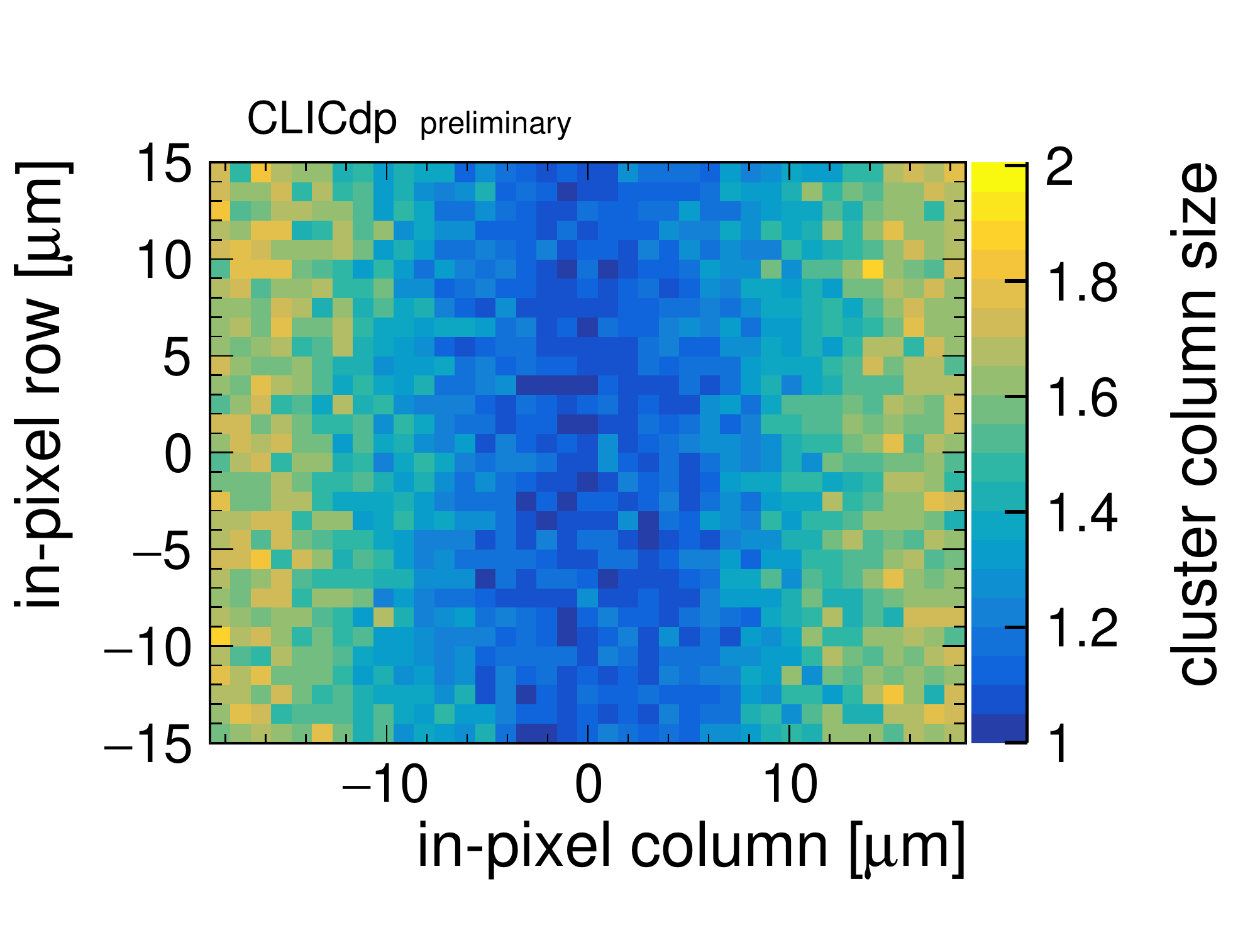}
	%	\caption{In-pixel cluster column size for an incident angle of 5\,deg.}
	%	\label{fig:in_pixel_5deg}
	\end{minipage}
	\hfill
	\begin{minipage}[b]{0.49\textwidth}
		\includegraphics[width=\linewidth]{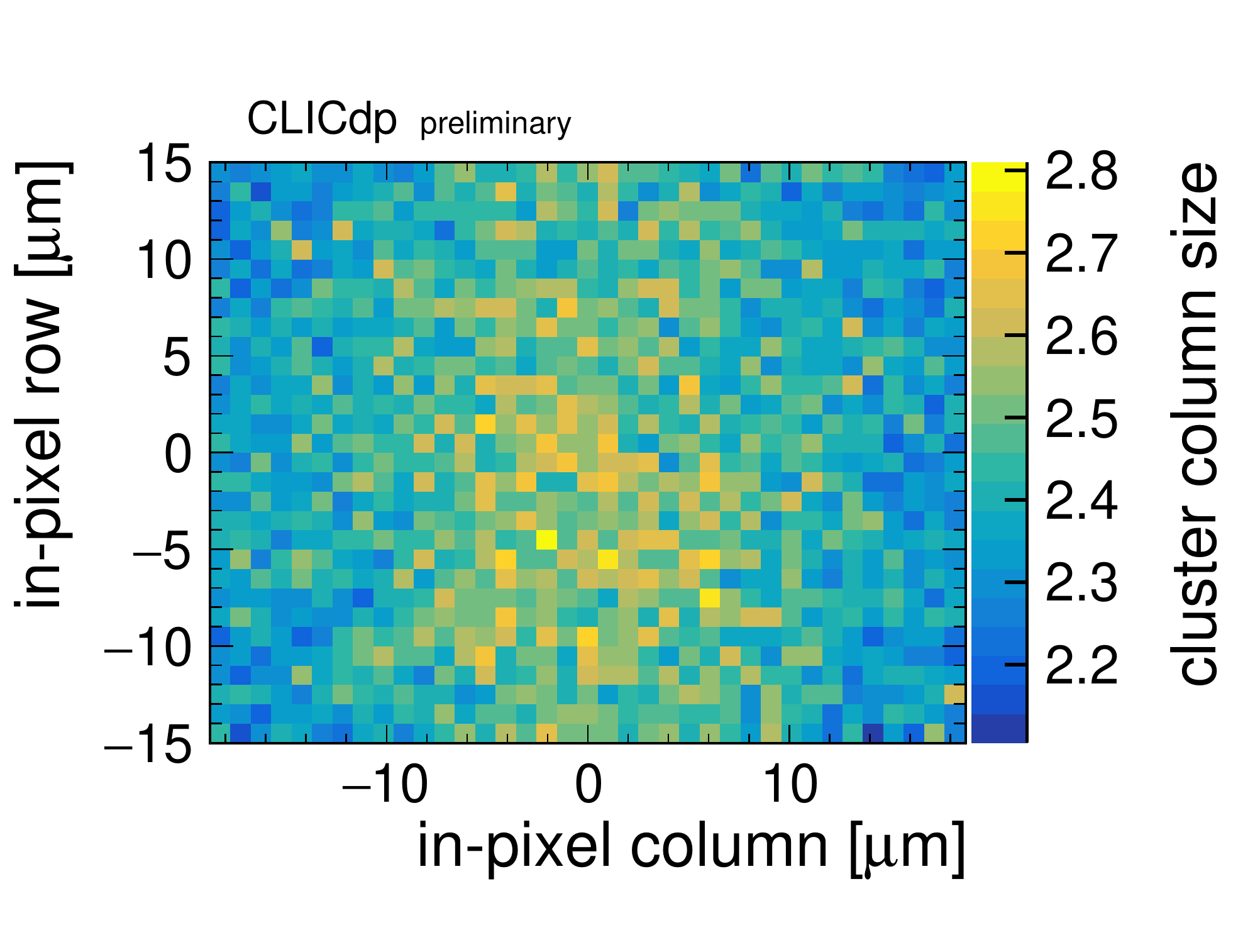}
	%	\caption{In-pixel cluster column size for an incident angle of 60\,deg.}
	%	\label{fig:in_pixel_60deg}
	\end{minipage}
	\caption{In-pixel cluster column size for an incident angle of $5^{\circ}$ (left) and $60^{\circ}$ (right) for an assembly with continuous n-layer.}
	\label{fig:in_pixel_cluster_size}
\end{figure}

%Contrary to the model assumptions, for CLICTD, the mean cluster size for perpendicular incident is greater than one.
%In the mean row cluster size distribution, shown in  Fig.~\ref{fig:cls_size}, a mean size of 1.49 at the minimum threshold can be observed owing to charge sharing via diffusion.
%The same holds true for the column cluster size, where a mean size of 1.40 was measured. 
%For small rotation angles, the diffusion effect plays a dominant role in the cluster size distribution, as illustrated in the in-pixel cluster column size in Fig.~\ref{fig:in_pixel_cluster_size}.
Contrary to the model assumptions, the diffusion effect plays a dominant role in the cluster size distribution for small rotation angles, as illustrated in the in-pixel cluster column size in Fig.~\ref{fig:in_pixel_cluster_size}.
In this representation, the cluster column size is plotted as a function of the track incident position in a pixel cell.  
On the left, the in-pixel cluster column size  at $\alpha = 5^{\circ}$ is depicted for an assembly with continuous n-layer.
For incident positions in the pixel centre, the size is close to one, as expected for no charge sharing.
In the pixel edges, the cluster size increases owing to charge sharing via diffusion. 
No significant rotation effects on the cluster size are observed at such a small angle. 

For higher incident angles, the rotation-induced charge sharing dominates, as shown on the right of Fig.~\ref{fig:in_pixel_cluster_size} for a rotation angle of $\alpha = 60^{\circ}$.
While particles that impinge on the pixel edges deposit most of their energy in the active region of a two pixels, an incident point in the pixel centre allows energy to be deposited in the active region of more than two cells, as sketched on the right of Fig.~\ref{fig:rotation_sketches}. 
It should be noted that the in-pixel track intercept position for inclined tracks does not correspond to the intersection of the track with the sensor surface.
In order to comply with the definition of the reconstructed cluster position, which is calculated with a charge-weighted centre-of-gravity algorithm, the intersection of the track with the sensor is in the central region of the active silicon traversed by the beam.
The exact location of the track intercept with the sensor is determined in the offline alignment procedure, where the detector position is reconstructed such that the spatial residuals between track position and reconstructed cluster position are minimized.
%For inclined tracks, the cluster position is calculated with a charge-weighted centre-of-gravity algorithm. 
%As a result, the cluster position is reconstructed in the central region of the active silicon traversed by the beam and the track position offline alignment is adapted accordingly.
In Fig.~\ref{fig:rotation_sketches}, the track position is approximated by a star. 
Consequently, the left track in the schematic is associated with a hit in the pixel edge and the right track in the pixel centre. 
 
\begin{figure}[tbh]
	\centering
	\includegraphics[width=0.5\textwidth]{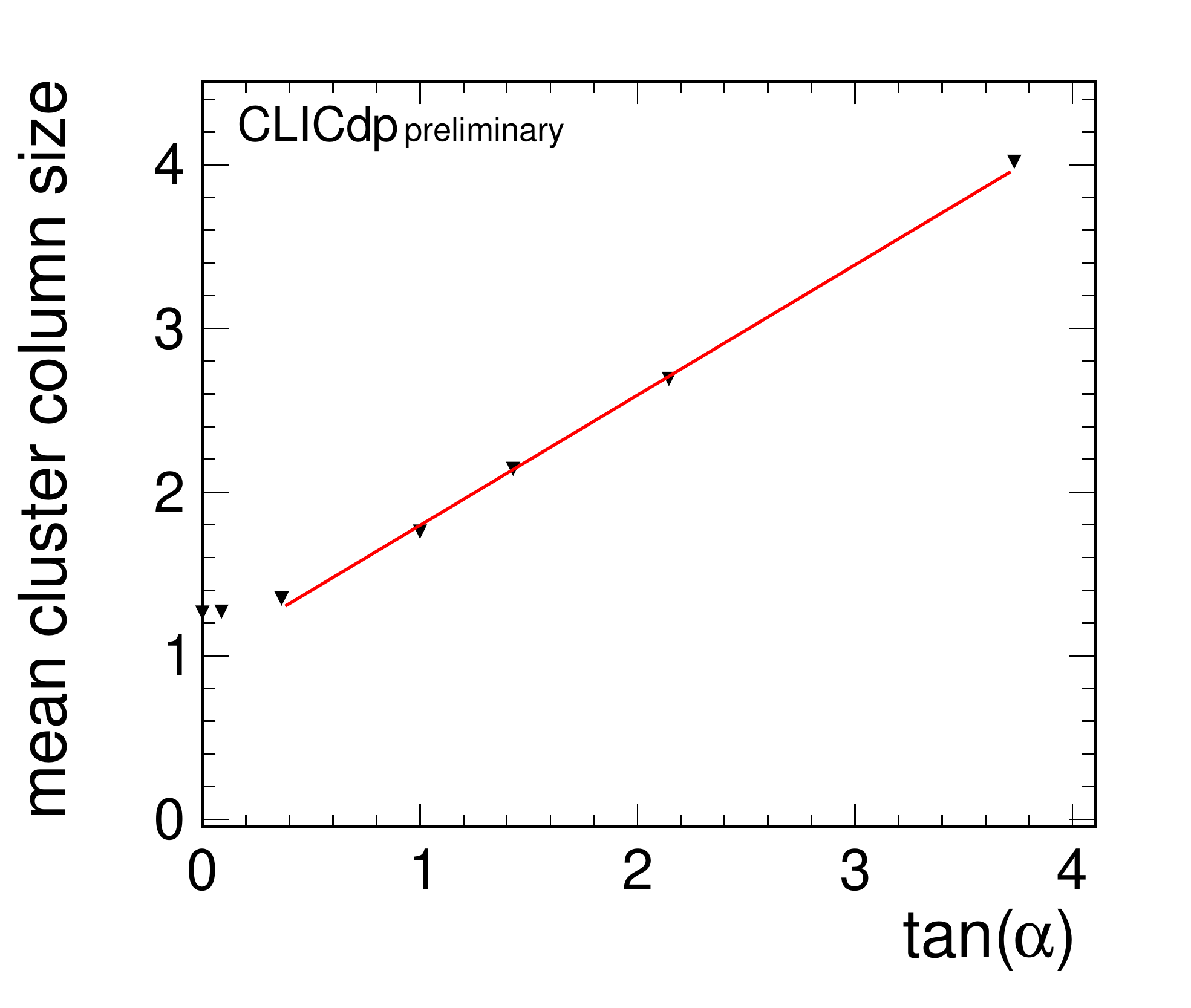}
	\caption{Mean cluster size in column direction as a function of the tangent of the rotation angle for an assembly with continuous n-layer.}
	\label{fig:active_depth_fit}
\end{figure}

%\begin{figure}[!tbp]
%	\centering
%	\begin{minipage}[b]{0.49\textwidth}
%		\includegraphics[width=\linewidth]{images/CLICTD_rotation_sketch.pdf}
%		\caption{Schematic representation of the CLICTD sensor for two different track incident angles. }
%		\label{fig:sketch_geometry}
%	\end{minipage}
%	\hfill
%	\begin{minipage}[b]{0.49\textwidth}
%		\includegraphics[width=\linewidth]{images/active_depth_fit_yaxis0.pdf}
%		\caption{Mean cluster size in column direction as a function of the rotation angle. }
%		\label{fig:active_depth_fit}
%	\end{minipage}
%\end{figure}
For the calculation of the active depth with Eq.~\ref{eq:active_depth}, the incident angles are required to be above 20$^{\circ}$ to ensure that the main contribution to the cluster size is the rotation-induced charge sharing.  
In Fig.~\ref{fig:active_depth_fit}, the mean cluster size in column direction as a function of the tangent of the incident angle is depicted for an assembly with continuous n-layer and \SI{50}{\micro m} thickness.
Eq.~\ref{eq:active_depth} is fitted to the curve, yielding an active depth of $d = (29.8 + 0.9 - 1.0)$\,\SI{}{\micro m}.
The uncertainty is estimated by refitting Eq.~\ref{eq:active_depth} with varied fit ranges.
The active depth $d$ corresponds to the thickness of the epitaxial layer.
From 3D TCAD simulations, the depletion depth is expected to be approximately \SI{23}{\micro m}, which implies that charge carriers created below the depletion zone can diffuse into the high electric field regions and contribute to the measured signal. 

%\begin{figure}[!tbp]
%	\centering
%	\begin{minipage}[b]{0.49\textwidth}
%		\includegraphics[width=\linewidth]{images/activeDepthFit_yaxis0.pdf}
%		\caption{Mean cluster size in column direction as a function of the tangent of the rotation angle for an assembly with continuous n-layer.}
%		\label{fig:active_depth_fit}
%	\end{minipage}
%	\hfill
%	\begin{minipage}[b]{0.49\textwidth}
%		\includegraphics[width=\linewidth]{images/efficiency_A1_A6_overlay_vertexConfErrorX.pdf}
%		\caption{Efficiency as a function of threshold for the process with continuous n-implant and different sensor thicknesses.}
%		\label{fig:efficiency_A1_A6_overlay}
%	\end{minipage}
%		%\caption{Efficiency as a function of threshold for the process with continuous n-implant (left) and gap in the n-implant (right) for different sensor thicknesses.}
%		%\label{fig:efficiency_overlay}
%\end{figure}

The analysis has been repeated with assemblies of both process types and with sensor thicknesses between \SI{300}{\micro m} and \SI{50}{\micro m}. 
The result for the active depth varied within the uncertainties, which indicates that the sensitive sensor depth is largely unaffected by the thinning process.

\section{Summary and Outlook}

Characterisation results of the monolithic CMOS sensor CLICTD in beam tests have been presented with focus on different operation voltages and inclined particle tracks.
It has been shown that a higher absolute bias voltage allows for operation at a lower detection threshold, which has an impact on the overall detector performance. 
The lower threshold improves the spatial resolution from \SI{4.8}{\micro m} at 200\,e$^-$ to \SI{4.6}{\micro m} at 140\,e$^-$.
Moreover, a timing resolution of \SI{5.8}{ns} has been achieved, together with a hit detection efficiency of $>$\SI{99.7}{\percent}. 

The active sensor depth has been determined with inclined particle tracks.
Regardless of the process variant and the total sensor thickness (in the range of \SI{300}{\micro m} to \SI{50}{\micro m}), an active depth of approximately \SI{30}{\micro m} has been found. 

It was shown that the CLIC tracking detector requirements are fulfilled in terms of timing and spatial resolution as well as efficiency and material budget.
The power consumption of CLICTD is also in line with the tracking detector requirements, as has been presented  elsewhere~\cite{clictd_design_characterization}.
Moreover, the results are relevant for future detector upgrades of e.g. the ALICE~\cite{alice_alpide} and ATLAS~\cite{atlas_mini_malta} experiment at CERN since CLICTD serves as a test-vehicle for studying the 180\,nm and 65\,nm CMOS processes chosen for the small collection electrode design. 

The characterisation of CLICTD assemblies fabricated on high resistivity Czochralski wafers, which allow for a thicker active layer, are foreseen for the future.
In addition, the test-beam results will be complemented with Monte Carlo-based simulation studies. 

%\begin{figure}[tbh]
%\includegraphics{fig01.eps}
%\caption{You can embed figures using the \texttt{\textbackslash includegraphics} command. Basically, figures should appear where they are cited in the text. You do not need to separate figures from the main text when you use \LaTeX\ for preparing your manuscript.}
%\label{f1}
%\end{figure}

\section*{Acknowledgements}
This work has been sponsored by the Wolfgang Gentner Programme of the German Federal Ministry of Education and Research (grant no. 05E15CHA).
The measurements leading to these results have been performed at the DESY II Test Beam Facility at DESY, Hamburg (Germany), a member of the Helmholtz Association (HGF).

%\appendix
%\newcommand{\newblock}{}
%\bibliographystyle{unsrt}	
%\bibliography{bibliography}

\end{document}